\documentclass[manuscript,format=acmlarge,screen,nonacm]{acmart}

\AtBeginDocument{%
  \providecommand\BibTeX{{%
    Bib\TeX}}}
\usepackage{hyperref}
\usepackage{caption}
\usepackage{subcaption}
\usepackage{lipsum}  
\usepackage[title]{appendix}
\usepackage[inkscapelatex=false]{svg}
\usepackage{todonotes}
\usepackage{tabularx}
\usepackage{mdframed}
\usepackage[normalem]{ulem} 
\usepackage{xcolor}         
\usepackage{multicol}

\newmdenv[linewidth=0pt,linecolor=white,innertopmargin=9,innerbottommargin=9,backgroundcolor=lightgray]{searchbox}

\newmdenv[linewidth=0pt,linecolor=white,innertopmargin=-3,innerbottommargin=6,backgroundcolor=lightgray,skipbelow=2pt,innerleftmargin=5,innerrightmargin=5,nobreak=true]{review}

\mdfsetup{skipabove=1em,skipbelow=1em}
\newmdenv[topline=false,rightline=false,bottomline=false,linewidth=4pt,linecolor=lightgray]{highlightbox}

\AtBeginDocument{%
  \providecommand\BibTeX{{%
    \normalfont B\kern-0.5em{\scshape i\kern-0.25em b}\kern-0.8em\TeX}}}

\def\@ACM@copyright@check@cc{}
\setcopyright{none}

\begin{document}



\author{Supraja Ramesh}
\email{supraja.ramesh@kit.edu}
\affiliation{%
  \institution{Karlsruhe Institute of Technology}
  \city{Karlsruhe}
  \country{Germany}
}
\orcid{0009-0007-3736-144X}

\author{Jonas Hummel}
\email{jonas.hummel@kit.edu}
\affiliation{%
  \institution{Karlsruhe Institute of Technology}
  \city{Karlsruhe}
  \country{Germany}
}
\orcid{0009-0005-8563-6175}

\author{Silvia Becker}
\affiliation{%
  \institution{Karlsruhe Institute of Technology, University Heart Center Freiburg-Bad Krozingen}
  \country{Germany}
}
\orcid{0000-0002-1135-6446}

\author{Christopher Clarke}
\affiliation{%
  \institution{University of Bath}
  \city{Bath}
  \country{United Kingdom}
}
\orcid{0000-0003-3916-7708}

\author{Jake Stuchbury-Wass}
\affiliation{%
  \institution{University of Cambridge}
  \city{Cambridge}
  \country{United Kingdom}
}
\orcid{0000-0001-6733-0504}

\author{Kai Kunze}
\affiliation{%
  \institution{Clausthal University of Technology}
  \city{Clausthal-Zellerfeld}
  \country{Germany}
}
\orcid{0000-0003-2294-3774}

\author{Axel Loewe}
\affiliation{%
  \institution{Karlsruhe Institute of Technology}
  \city{Karlsruhe}
  \country{Germany}
}
\orcid{0000-0002-2487-4744}

\author{Michael Beigl}
\affiliation{%
  \institution{Karlsruhe Institute of Technology}
  \city{Karlsruhe}
  \country{Germany}
}
\email{michael.beigl@kit.edu}
\orcid{0000-0001-5009-2327}

\author{Tobias R{\"o}ddiger}
\affiliation{%
  \institution{Karlsruhe Institute of Technology}
  \city{Karlsruhe}
  \country{Germany}
}
\email{tobias.roeddiger@kit.edu}
\orcid{0000-0002-4718-9280}

\renewcommand{\shortauthors}{Ramesh et al.}
\newcommand{\participants}{330}
\newcommand{\androidParticipants}{165}
\newcommand{\iPhoneParticipants}{90}
\newcommand{\totalIphone}{165}

\newcommand{\male}{181}
\newcommand{\female}{143}
\newcommand{\diverse}{6}

\newcommand{\ageMin}{18}
\newcommand{\ageMax}{67}
\newcommand{\ageMean}{28.69}
\newcommand{\ageStd}{11.93}

\newcommand{\germanGroup}{88.18}
\newcommand{\nonGermanGroup}{11.82}

\newcommand{\totalSessions}{1,530,950}
\newcommand{\medianDuration}{0.7}
\newcommand{\qOneDuration}{0.1}
\newcommand{\qThreeDuration}{5.7}

\newcommand{\medianVolume}{64.84}
\newcommand{\qOneVolume}{56.05}
\newcommand{\qThreeVolume}{73.00}

\newcommand{\mainAlpha}{0.05}

\newcommand{\iPhoneVsAndroidAlpha}{0.005}
\newcommand{\pldMusic}{45.15}
\newcommand{\pldMovie}{41.52}
\newcommand{\pldPodcast}{38.48}
\newcommand{\pldCall}{35.45}
\newcommand{\pldGaming}{14.85}
\newcommand{\pldOthers}{3.64}

\newcommand{\ancActive}{52.12}
\newcommand{\ancDidNot}{28.18}
\newcommand{\ancNoSupport}{13.33}
\newcommand{\ancNotAware}{4.85}
\newcommand{\totalANCUseParticipants}{221}

\newcommand{\pldCommuting}{48.79}
\newcommand{\pldSport}{43.03}
\newcommand{\pldWork}{33.64}
\newcommand{\pldHome}{30.91}
\newcommand{\pldWalkNature}{24.85}

\newcommand{\ancWork}{57.92}
\newcommand{\ancSport}{36.65}
\newcommand{\ancCommuting}{35.75}
\newcommand{\weekendObservations}{356}
\newcommand{\jointB}{0.18}
\newcommand{\jointZ}{4.17}
\newcommand{\jointp}{0.001}

\newcommand{\sessionB}{-0.16}
\newcommand{\sessionZ}{-2.16}
\newcommand{\sessionp}{0.03}

\newcommand{\durationB}{-0.09}
\newcommand{\durationZ}{-1.22}
\newcommand{\durationP}{0.22}

\newcommand{\yearTotalParticipants}{88}
\newcommand{\yearTotalObservations}{335}

\newcommand{\yearTwentyM}{37.17}
\newcommand{\yearTwentyMAbstract}{37}

\newcommand{\yearTwentyOneM}{48.45}
\newcommand{\yearTwentyOneb}{11.28}
\newcommand{\yearTwentyOnez}{1.38}
\newcommand{\yearTwentyOnep}{0.17}

\newcommand{\yearTwentyTwoM}{63.39}
\newcommand{\yearTwentyTwob}{26.22}
\newcommand{\yearTwentyTwoz}{2.28}
\newcommand{\yearTwentyTwop}{0.02}

\newcommand{\yearTwentyThreeM}{56.52}
\newcommand{\yearTwentyThreeb}{19.35}
\newcommand{\yearTwentyThreez}{2.62}
\newcommand{\yearTwentyThreep}{0.009}

\newcommand{\yearTwentyFourM}{63.68}
\newcommand{\yearTwentyFourb}{26.51}
\newcommand{\yearTwentyFourz}{3.17}
\newcommand{\yearTwentyFourp}{0.002}
\newcommand{\yearTwentyFourMAbstract}{64}

\newcommand{\psyBonferroni}{0.003}
\newcommand{\psySocialBonferroni}{0.02}

\newcommand{\medExpSeek}{3.5}
\newcommand{\medThrill}{2.5}
\newcommand{\medBoredom}{3.0}
\newcommand{\medDisinhibition}{2.5}
\newcommand{\medOverall}{3.0}
\newcommand{\overallMusicTime}{0.28}
\newcommand{\overallMusicTimep}{0.0001}

\newcommand{\overallPLDUsage}{0.17}
\newcommand{\overallPLDUsagep}{0.002}

\newcommand{\disinhibitionPLDFreq}{0.20}
\newcommand{\disinhibitionPLDFreqp}{0.0001}

\newcommand{\disinhibitionPLDUsage}{0.25}
\newcommand{\disinhibitionPLDUsagep}{0.0001}

\newcommand{\disinhibitionMusic}{0.33}
\newcommand{\disinhibitionMusicp}{0.0001}

\newcommand{\thrillPLDUsage}{0.22}
\newcommand{\thrillPLDUsagep}{0.0001}

\newcommand{\thrillMusic}{0.30}
\newcommand{\thrillMusicp}{0.0001}

\newcommand{\medOpenness}{3.5}
\newcommand{\medConscientiousness}{3.5}
\newcommand{\medAgreeableness}{3.5}
\newcommand{\medExtraversion}{3.5}
\newcommand{\medNeuroticism}{3.0}
\newcommand{\opennessPLDFreq}{0.13}
\newcommand{\totalSocialAvoidance}{91.82}
\newcommand{\completeAvoidance}{57.88}
\newcommand{\mostlyAvoidance}{33.94}
\newcommand{\agreedIncreasedShielding}{57.27}
\newcommand{\disagreedIncreasedShielding}{25.76}
\newcommand{\polarizationShielding}{21.52}
\newcommand{\agreedIncreasedIsolation}{49.09}
\newcommand{\disagreedIncreasedIsolation}{31.52}
\newcommand{\polarizationIsolation}{24.55}
\newcommand{\shieldVsIsolation}{0.62}
\newcommand{\shieldVsIsolationp}{0.001}

\newcommand{\isolationVsSocial}{-0.16}
\newcommand{\isolationVsSocialp}{0.003}

\newcommand{\shieldVsSocial}{-0.05}
\newcommand{\shieldVsSocialp}{0.34}

\newcommand{\demographicsBonferroni}{0.01}

\newcommand{\maleNumber}{49}
\newcommand{\femaleNumber}{41}

\newcommand{\maleMedUsageTime}{22.20}
\newcommand{\femaleMedUsageTime}{26.40}
\newcommand{\genderUsageTimeU}{830}
\newcommand{\genderUsageTimeP}{0.16}
\newcommand{\genderUsageTimeR}{0.15}

\newcommand{\maleMedUsageVolume}{62.32}
\newcommand{\femaleMedUsageVolume}{60.84}
\newcommand{\genderUsageVolumeU}{1044}
\newcommand{\genderUsageVolumeP}{0.75}
\newcommand{\genderUsageVolumeR}{0.03}

\newcommand{\UnderCount}{51}
\newcommand{\MidCount}{22}
\newcommand{\OldCount}{17}

\newcommand{\UnderMedUsageTime}{24.60}
\newcommand{\MidMedUsageTime}{22.20}
\newcommand{\OldMedUsageTime}{26.40}
\newcommand{\ageUsageTimeH}{0.45}
\newcommand{\ageUsageTimeP}{0.80}

\newcommand{\UnderMedVolume}{64.25}
\newcommand{\MidMedVolume}{58.81}
\newcommand{\OldMedVolume}{62.74}
\newcommand{\ageVolumeH}{8.95}
\newcommand{\ageVolumeP}{0.01}

\newcommand{\ageVolumeUnderMidU}{818}
\newcommand{\ageVolumeUnderMidP}{0.002}
\newcommand{\ageVolumeUnderMidR}{0.36}
\newcommand{\volumeObservations}{13718}
\newcommand{\totalParticipantsVolume}{88}

\newcommand{\volumeTwenty}{67.48}

\newcommand{\volumeTwentyOne}{64.61}
\newcommand{\volumeTwentyOneZ}{-1.97}
\newcommand{\volumeTwentyOneb}{-2.87}
\newcommand{\volumeTwentyOnep}{0.05}

\newcommand{\volumeTwentyTwo}{62.40}
\newcommand{\volumeTwentyTwoZ}{-2.54}
\newcommand{\volumeTwentyTwob}{-5.08}
\newcommand{\volumeTwentyTwop}{0.01}
\newcommand{\volumeYearDiff}{5}

\newcommand{\volumeTwentyThree}{64.36}
\newcommand{\volumeTwentyThreeZ}{-1.45}
\newcommand{\volumeTwentyThreeb}{-3.12}
\newcommand{\volumeTwentyThreep}{0.15}

\newcommand{\volumeTwentyFour}{63.98}
\newcommand{\volumeTwentyFourZ}{-1.44}
\newcommand{\volumeTwentyFourb}{-3.50}
\newcommand{\volumeTwentyFourp}{0.15}

\newcommand{\medEightyOneToNinety}{2.57}
\newcommand{\medNinetyOneToHundred}{0.39}
\newcommand{\medOverHundred}{0}
\newcommand{\partOverHundred}{8.89}
\newcommand{\partOverNinety}{9}

\newcommand{\participantWeekExceed}{4.01}
\newcommand{\participantAtleastOneWeekExceed}{35}
\newcommand{\zeroDayRate}{48.09}


\begin{abstract}
Ear-worn devices are evolving from audio-playback tools into sensing platforms for health, interaction, and context-awareness. Yet, earable systems are typically designed and evaluated under strong assumptions about how long, how often, and in which situations people actually wear personal listening devices (PLDs). To ground these assumptions in-the-wild behavior, we combine a survey of \participants{} adults with multi-year, passively logged headphone audio-exposure records donated via Apple Health by \iPhoneParticipants{} of them. We characterize where and when people use PLDs, how logged use has changed in recent years, and how psychological traits and social context associate with PLD usage. Our results show that logged mean daily use has increased from 37 minutes in 2020 to 64 minutes in 2024. Listening was intermittent: no listening was logged on 48\,\% of participant-days in 2024, and sessions were fewer but longer on weekends. Sensation seeking, particularly disinhibition, showed small to medium positive associations with self-reported PLD use. Younger adults listened at higher volumes than the 25-34 group. High-volume exposure was uncommon, with only 4\,\% of participant-weeks exceeding World Health Organization (WHO) safe-listening limits. Finally, most participants also reported avoiding PLD use in social situations. We translate these findings into implications for earable computing: realistic expectations of intermittent rather than continuous wear, contextual coverage that anticipates systematic gaps, targeted safe-listening interventions, and personalization grounded in psychosocial and demographic profiles rather than assumptions of uniform use.
\end{abstract}

\title{Characterizing In-the-Wild Personal Listening Device Use to Inform Earable Application Design}

\maketitle





\section{Introduction}
\label{sec:introduction}

Personal listening devices (PLDs), such as headphones and earphones, are ubiquitous in everyday life, with hundreds of millions of devices in circulation worldwide~\cite{statista2025headphones}. They are among the most widely adopted wearables, shaping how people engage with media, manage attention, and interact with their environments. More recently, PLDs have been augmented with additional sensing functionality and a new subset known as ``earables''~\cite{roddiger2022sensing} (also known as ``hearables''~\cite{goverdovsky2017hearables}) have emerged with applications in fitness tracking~\cite{stromback2020mm}, continuous vitals monitoring~\cite{zhang2025continuous,roddiger2019towards}, disease management~\cite{aziz2025unobtrusive,choi2022health}, safety detection~\cite{kaveh2024wireless}, and wellbeing support~\cite{min2018cross,hu2024breathpro,mandekar2021earable}. Earables leverage the ubiquity of PLDs as an existing socially accepted platform, with research papers using the popularity of PLDs as a basis for future earable adoption in various application domains~\cite{roddiger2022sensing}.
However, longitudinal in-the-wild research into earables has been highlighted as a significant shortcoming of existing earable research~\cite{roddiger2022sensing}. A deeper understanding of how PLDs are currently used on an everyday basis would provide a foundation for future earable research to build upon, but this is currently lacking~\cite{reinelt2016design}. We address this by exploring how PLDs are used in the wild, revealing what factors need to be taken into consideration when deploying ear-based devices on a long-term basis.

It is well-established that PLDs are used to listen to music, audio books, or podcasts when commuting~\cite{liikkanen2010observing}, studying~\cite{alshaikh2025association}, or exercising~\cite{hallett2017music}. 
Earables have the potential to be deployed in a much larger range of context-specific applications, but to date, most of these systems have been evaluated in controlled laboratory or short-term pilot settings, and typically presume that users will wear devices for extended periods across different everyday contexts. There is little empirical evidence on how people actually adopt and wear PLDs in the wild over time, raising questions about whether or not people are likely to wear devices over longer periods. Understanding how PLDs are currently used is an important first step for understanding realistic expectations about wear time, adherence, and contextual coverage, which can be used to inform long-term deployments of earables. To better understand these research gaps, we address the following use-case focused research questions: 

\begin{enumerate}
    \item[\emph{RQ1:}] What are the self-reported everyday contexts of PLD and Active Noise Cancellation (ANC) use?
    \item[\emph{RQ2:}] How are PLDs used over time?
\end{enumerate}

PLDs are used across broad and diverse populations. Prior research has shown how different user groups engage with PLDs in distinct ways: some favor short, intermittent listening, others use them for long, focused sessions~\cite{kim2021analysis}, and some use them to avoid social interactions~\cite{harshitha2017survey}. In addition, studies have associated PLD usage with demographic factors such as age or gender~\cite{hodgetts2007effects, torre2008young} and with psychological dimensions including anxiety~\cite{ekcsi2019headphone}, depression~\cite{vzivojinovic2023personal}, and sensation seeking~\cite{kallinen2007comparing}. This suggests that a person's demographic characteristics, as well as their psychological traits and states, relate to PLD usage behavior. However, prior studies relied on self-reports and controlled studies to examine the demographic factors associated with PLD usage, and these results were not evaluated with objective usage data. Furthermore, the relationship between facets of sensation seeking and personality and PLD usage habits, as well as the social contexts of PLD usage, has not yet been analyzed. In addition, safety is a critical consideration for long-term PLD use, and extended listening at high volumes can cause ear pain~\cite{harshitha2017survey}, psychological discomfort~\cite{vzivojinovic2023personal}, and noise-induced hearing loss~\cite{lee2021personal}. The World Health Organization (WHO) emphasizes that both sound level and listening duration jointly determine risk, recommending that leisure noise exposure remains below approximately 80 dB(A) for a maximum of 40 hours per week~\cite{WHO2019SafeListening}. However, there is little in-the-wild, longitudinal evidence on how loud and how long people actually listen in everyday life, which has implications for proposing earable-based solutions that may exacerbate these issues. We address these with three user-focused research questions: 
\begin{enumerate}
    \item[\emph{RQ3:}] How are psychological traits and social contexts associated with PLD use?
    \item [\emph{RQ4:}] How do demographics relate to PLD usage?
    \item [\emph{RQ5:}] How loud are people listening, and is it within WHO guidelines?
\end{enumerate}

We address these research questions using a multi-source study that combines self-report and passively logged behavioral data to understand PLD usage in everyday life. We recruited \participants{} participants who completed a survey capturing demographics, personality traits, social and contextual use cases, and self-reported listening habits (RQ1, RQ3, RQ4). A subset of iPhone users ($n$ = \iPhoneParticipants{}) additionally shared their historical PLD usage data, automatically recorded in Apple Health. Linking survey responses with longitudinal, device-logged behavior allows us to quantify in-the-wild PLD usage patterns, characterize temporal dynamics (RQ2), and estimate listening exposure relevant to hearing health guidelines (RQ5). \autoref{tab:rqs} summarizes these questions and the respective data sources to answer them.  

This paper makes three contributions to HCI and earable research: (i) we characterize temporal patterns of PLD use in natural settings over five years; (ii) we analyze how activities, demographics, psychological traits, and social contexts relate to PLD usage, revealing individual differences that relate to adoption and practices;  (iii) we assess safe listening by comparing in-the-wild exposure levels against WHO safe listening guidelines, examining the distribution of sessions across volume categories and weekly sound exposure. Collectively, these contributions deepen our understanding of everyday PLD practices, challenge many of the assumptions that commonly inform earable design, and provide a realistic PLD usage trend for grounded earable research and informing methodological design of in-the-wild longitudinal research studies.

\begin{table}[!b]
\begin{center}
\small
\caption{Research questions investigated in this study are organized into five thematic categories. Each question was addressed using either survey data, historical usage data, or a combination of both, as indicated in the second column. Participant numbers varied according to the available data source, as shown in the last column.}
\label{tab:rqs}
\begin{tabularx}{\textwidth}{Xll} 

 \textbf{Research Question} & \textbf{Source} & \textbf{Participants} \\
 \toprule
 \textbf{Use Cases} & & \\
 (RQ1) \hyperlink{RQ_I}{What are the self-reported everyday contexts of PLD and ANC use?} & Survey & \participants{} \\
 (RQ2) \hyperlink{RQ_II}{How are PLDs used over time?} & Historical Data & \iPhoneParticipants{} \\
 \midrule
 \textbf{Users} & & \\
 (RQ3) \hyperlink{RQ_III}{How are psychological traits and social contexts associated with PLD use?} & Survey & \participants{}  \\
 (RQ4) \hyperlink{RQ_IV}{How do demographics relate to PLD usage?} & Survey, Historical Data & \iPhoneParticipants{} \\
 (RQ5) \hyperlink{RQ_V}{How loud are people listening, and is it within WHO guidelines?} & Historical Data & \iPhoneParticipants{} \\
 \midrule
\end{tabularx}
\end{center}

\end{table}
\section{Related Work}
\label{sec:related_work}
Understanding how, when, and why people use personal listening devices in everyday life has implications for earable research and applications. Prior research has established demographic trends, psychosocial functions, and health risks associated with PLD use, but relies predominantly on self-reported measures that have already been shown to be inaccurate \cite{paping2021smartphone}. In this section, we review the main strands of PLD research to establish gaps that longitudinal, passively logged data can address.

\subsection{Constraints of Prior PLD Research}
\label{subsec:method_limitations}
Past research on PLD usage has been based primarily on two methodological approaches: self-reports~\cite{fligor2014cultural, torre2008young, alomari2024cultural, lee2021personal, ekcsi2019headphone, harshitha2017survey} and controlled experiments~\cite{torre2008young, hohneck2024hemodynamic, belyad2024effect}. These approaches have established basic demographics, usage patterns, and health risks, but they are limited by recall bias, subjective judgments, and artificial settings, which raises questions about how well they reflect everyday PLD behavior~\cite{portnuff2011teenage,alkharabsheh2025personal}. Recent work by~\citet{neitzel2022toward} introduced passive data collection via Apple's Health and Research apps, analyzing 121,010 participants with 21,128,363 hours and 12,342,758 participant-days of data and showed that only 10\% of participant-days exceeded WHO safe-listening limits~\cite{WHO2019SafeListening}. However, this study primarily described the methodology, reported high-level preliminary results, and focused on sound exposure without examining temporal patterns, demographic differences, or psychosocial correlates.

Building on this passive data collection approach, the present work analyzes longitudinal, passively logged PLD data combined with survey measures to characterize in-the-wild usage behavior across multiple dimensions relevant to earable design and hearing health.

\subsection{PLD Usage Patterns}
Context and user characteristics relate to PLD behavior, but findings remain inconsistent. For example, men have been reported to listen at higher levels and for longer durations~\cite{torre2008young}, although other work finds no gender differences~\cite{fligor2014cultural}. Similarly, some studies show that noisy environments prompt listeners to increase playback levels by 6–10 dB~\cite{hodgetts2007effects}, while others report no location-based differences~\cite{fligor2014cultural}. These inconsistencies highlight that PLD use is heterogeneous and sensitive to both situational and individual factors that cannot be perfectly captured through point measures.

Work on ANC has highlighted a wide variety of use cases, including sleep, travel, hospital care, and open-plan offices, and suggests that ANC can change both experience and listening behavior~\cite{bhatia2025revolutionizing,mueller2022using}. For example, studies in noisy environments indicate that PLDs with noise reduction led users to select lower listening levels, while survey work shows strong user demand for better noise cancellation in everyday PLDs, positioning ANC as a key feature in contemporary PLD choice~\cite{hodgetts2007effects,seol2022influence,rane2022survey}. However, most of this research captures adoption and preferences for specific applications, leaving open how often, how long, and in which everyday situations people actually activate ANC in their routine PLD use.

To provide an ecologically valid perspective, the present study links longitudinal, passively logged PLD behavior with demographic data, offering an in-the-wild account of usage patterns that can inform in-the-wild PLD and earable study design. Further, we evaluate ANC adoption and characterize the contexts in which users rely on ANC in daily life through self-reports.

\subsection{Psychosocial Factors in PLD Usage}
Beyond demographics, PLDs serve psychosocial functions such as managing mood, regulating arousal, and shaping social interactions~\cite{ekcsi2019headphone,harshitha2017survey}. Survey studies have associated higher anxiety with more frequent PLD use in certain situations, and users often employ PLDs to avoid social contact, structure daily routines, or cope with stress, indicating that PLDs act as tools for both withdrawal and self-management~\cite{ekcsi2019headphone,harshitha2017survey,salami202514}. Research with adolescents further links extended daily PLD use to elevated depressive symptoms and to more negative attitudes toward noise, and work on sensation seeking suggests that people with certain personality traits show more physiological arousal to PLD use than others~\cite{vzivojinovic2023personal,kallinen2007comparing}.

These studies establish that psychosocial factors are associated with PLD use. To build on this work, we examined how social factors and psychological traits (personality and sensation seeking) relate to PLD usage. 

\subsection{Health Risks Associated with PLD Use}
Clinical research has established that high recreational sound exposure, including from PLDs, contributes to tinnitus, self-reported hearing difficulties, and early signs of hearing loss, particularly among adolescents and young adults~\cite{clark1992hearing,jokitulppo1997estimated,ramage2019tinnitus}. Surveys in school and online populations report early onset of PLD use, substantial proportions of users listening at high or maximum volume, frequent reports of hearing-related symptoms, and low uptake of hearing protection despite awareness of risks~\cite{lee2021personal,alomari2024cultural}. These studies underscore PLDs as a meaningful source of leisure noise exposure, yet almost all quantify exposure based on self-reported duration and volume, which are vulnerable to recall error and subjective interpretation of ``loudness''~\cite{portnuff2011teenage,alkharabsheh2025personal}.

By leveraging historical PLD usage data captured passively in everyday conditions, the present study offers a complementary, behaviorally grounded view on typical exposure levels and patterns that can support more realistic risk estimates and inform the design of timely, data-driven interventions, for instance in line with WHO--ITU safe listening recommendations~\cite{WHO2019SafeListening}.

\subsection{Earable Computing and the Usage Assumption Gap}
Earable computing has emerged alongside the rapid adoption of True Wireless Stereo earbuds with ANC~\cite{fan2023design}. Earable systems have been proposed for mobile health~\cite{zhang2025continuous, aziz2025unobtrusive, choi2022health}, hands-free interaction techniques~\cite{roddiger2021earrumble,matthies2013inear,iguma2023input}, and context-aware computing~\cite{lyu2024earda, franklin2021designing}, positioning ear-worn devices as a promising next-generation platform for mobile computing and HCI. However, these earable applications often assume that users wear such devices consistently over extended periods, although the empirical basis for these assumptions is thin, with most earable papers relying on short-term deployments. Hence, there is a lack of empirical evidence on in-the-wild PLD usage behavior. Understanding this will help build more context-specific applications with practical relevance. For instance, understanding attitudes towards using PLDs in social situations will help explain the adoption of earable applications such as conversational wellbeing~\cite{min2018cross}, while understanding the temporal patterns of usage will reveal when people habitually wear PLDs, which could help harness these periods for effective in-field data collection.

This paper provides longitudinal evidence on temporal trends in PLD use and attitudes toward wearing PLDs in social situations, thereby grounding earable design in realistic patterns of when and how users actually wear devices.

\section{Methodology}
\label{sec:methodology}
Prior research has relied mainly on self-reports or controlled laboratory studies and has rarely explored PLD usage in naturalistic contexts. To bridge this gap, we adopted a multi-source approach combining longitudinal, passively logged data with self-reported survey responses. We conducted the study with adults aged 18 years and older, combining survey responses from all participants with the option for iPhone users to donate historical PLD usage and health data via HealthKit. The study followed the university’s ethical guidelines and the Declaration of Helsinki and received formal approval from the ethics committee (Application Number: A2025-014). Participants were recruited through campus events, LinkedIn professional networks, and social media outreach over a period of 8 months. Participants at the campus events received a small refreshment, while no compensation was provided to participants recruited through online platforms.

\begin{figure}[!t]
    \includegraphics[width=\textwidth]{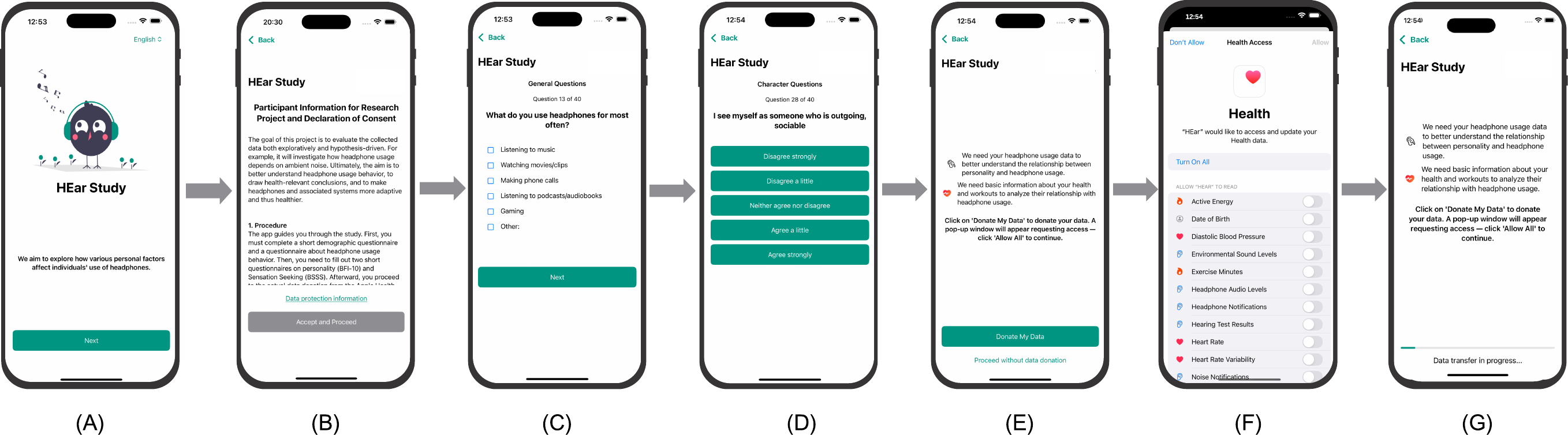}
    \caption{Steps of the study app for participants with an iPhone available via Apple's App Store. (A) The app welcomes the user and guides them through each step; (B) Participants are presented with a consent form that informs them about the study; (C) Participants answer a set of survey questions on demographics, PLD usage; and (D) psychology questionnaires; (E) Access to HealthKit data is described and (F) participants grant access to the HealthKit metrics collected in the study; (G) the data is extracted and transferred to the server.}
    \Description
    [Study app workflow for iPhone participants.] {
    The figure shows a sequence of seven smartphone screenshots (A–G) connected by arrows, illustrating the study app’s onboarding process.
    (A) The welcome screen introduces the HEar Study and invites participants to begin.
    (B) A consent form explains the study procedures and requires agreement to proceed.
    (C) The app displays survey questions about demographics and PLD usage.
    (D) Psychology-related questionnaires are presented.
    (E) A screen describes the request for HealthKit data access.
    (F) A native iOS Health permissions screen lists individual data types (e.g., audio exposure, exercise, heart rate) for participants to grant access.
    (G) The final screen confirms that data will be extracted and securely uploaded to the study server.
    }
  \label{fig:screen}
\end{figure}

\subsection{Study Procedure}
Participants were invited to join the study by scanning a QR code with their smartphones. Upon scanning, iPhone users were redirected to the Apple App Store to download the study app, which facilitated both the questionnaire and data donation processes. In contrast, non-iPhone users were redirected to the Sosci Survey platform solely to complete the questionnaire, as they could not participate in data donation. The survey and the app were available in both German and English. All participants first completed a consent declaration outlining the age eligibility requirement and emphasizing the anonymous nature of the study, with no mechanism existing to trace or remove donated data once submitted, ensuring complete participant anonymity. After giving informed consent, participants then completed a questionnaire covering demographic information and PLD usage habits, followed by personality assessment questions. iPhone users were subsequently guided to the data donation page. Participants could either choose to donate data or proceed without data donation (\autoref{fig:screen} (E)). If they chose to donate data, they could review and selectively authorize the sharing of specific health metrics (\autoref{fig:screen} (F)). Upon authorization, the app queried the selected data and transmitted it to a secure server, where it was stored alongside the survey responses. The survey took approximately 3 minutes to complete. 

\subsection{Measures}
Our study combines self-reports via questionnaires and a custom smartphone app to extract automatically and passively logged historical data.

\subsubsection{Questionnaire}

\begin{table}[!t]
\footnotesize
\caption{Survey questions with their corresponding response options. Items marked with * allowed multiple responses, and the ``Other'' option required a free-text response. The lines separate different question categories: demography, PLD usage, and psychological aspects.}
\centering 
\begin{tabularx}{\textwidth}{lX}

\textbf{Question} & \textbf{Options} \\
\toprule
1. Country of Residence & Dropdown with all the country names\\
2. \hypertarget{q2}{Age} & Text box (accepts only numbers $\geq$ 18) \\
3. \hypertarget{q3}{Gender} & Male, Female, Diverse \\
\midrule
4. \hypertarget{q4}{How often do you use headphones?} & Daily, Several times a week, Once a week, Less than once a week, Never \\
5. \hypertarget{q5}{What do you use headphones for most often?*} & Music, Movies/clips, Phone calls, Podcasts/audiobooks, Gaming, Other:\_\_\_ \\
6. \hypertarget{q6}{In which situations do you use headphones most often?*} & Working/studying, Sports/training, Commuting, At home, Walking/nature, Other:\_\_\_ \\
7. \hypertarget{q7}{How much time per week do you use headphones?} & Less than 1 hour, 1–5 hours, 6–10 hours, More than 10 hours \\
8. \hypertarget{q8}{Do you use noise-canceling/ANC features?} & Almost always, Frequently, Occasionally, No, No my headphones do not have ANC functionality, I don't know \\
9. \hypertarget{q9}{If yes, in which situations do you use ANC?*} & Working/studying, Sports/training, Commuting, At home, Walking/nature, Other:\_\_\_ \\
10. \hypertarget{q10}{How many hours per week for music with headphones?} & Less than 1 hour, 1–5 hours, 6–10 hours, More than 10 hours \\
11. \hypertarget{q11}{How do you feel about using headphones in social situations?} & Avoid, Occasionally, Often \\
12. \hypertarget{q12}{Do you feel socially isolated by headphones?} & Disagree strongly, Disagree a little, Neutral, Agree a little, Agree strongly \\
13. \hypertarget{q13}{Do you feel socially shielded by headphones?} & Disagree strongly, Disagree a little, Neutral, Agree a little, Agree strongly \\
\midrule
14. - 23. BFI-10 Questionnaire~\cite{rammstedt2013short} & Evaluates: Extraversion, Agreeableness, Conscientiousness, Neuroticism, and Openness\\
24. - 31. BSSS Questionnaire~\cite{hoyle2002reliability} &  Evaluates: Experience Seeking, Boredom Susceptibility, Thrill and Adventure Seeking, Disinhibition, and overall sensation seeking tendency\\
\end{tabularx}

\label{tab:survey}
\end{table}

The questionnaire administered in this study comprised four key sections: demographics, general PLD usage preferences, the 10-Item Big Five Inventory (BFI-10) personality questionnaire~\cite{rammstedt2013short}, and the Brief Sensation Seeking Scale (BSSS)~\cite{hoyle2002reliability} questionnaire.
An overview of the questionnaire is given in \autoref{tab:survey}. Demographic questions addressed factors such as age and gender, which could relate to PLD usage behaviors. Questions regarding PLD usage duration and purpose were included to provide a high-level overview of listening behaviors. The last two sections assessed personality traits to explore their potential correlations with PLD usage. 

The BFI-10 questionnaire consists of ten items, with two items corresponding to each of the five personality dimensions: Extraversion (sociability and assertiveness), Agreeableness (altruism and interpersonal trust), Conscientiousness (goal-directed and disciplined behavior), Neuroticism (emotional instability and tendency toward anxiety), and Openness (intellectual curiosity and interest in new experiences). Each trait was measured using two items out of the 10-item questionnaire, with specific items requiring reverse scoring. Trait scores were calculated by averaging across the two items per dimension.

The BSSS questionnaire consists of eight items assessing an individual's sensation seeking tendencies across four dimensions: Experience Seeking (seeking novel experiences through mind and senses), Boredom Susceptibility (aversion to repetitive experiences), Thrill and Adventure Seeking (desire for risky physical activities), and Disinhibition (seeking sensation through social disinhibition). Each facet was assessed using two items from the 8-item questionnaire. Facet scores were averaged across the two items per dimension, and an overall BSSS score (general tendency to sensation seeking) was computed by averaging all eight items.

\subsubsection{Mobile App}
To extract historical data from the user's phone, we implemented a native iOS application for this study. Within the app, users first complete the questionnaire described above and then authorize the donation of historical data. The Health app on iPhone devices tracks, consolidates, and stores a wide range of health metrics and PLD usage information collected directly from the device and through third-party applications and accessories such as smartwatches and PLDs. Apple has developed the HealthKit API to enable third-party developers to securely add and read health data, encouraging the creation of customized applications. Health data is collected upon user consent from the day a person begins using an iPhone and can be backed up to iCloud, ensuring continuity even if the device is changed. The longitudinal nature of this data enables retrospective analysis of hearing habits over time, particularly among individuals who have used iPhones for extended periods. 

During the data donation step, we queried the HealthKit API to retrieve the Headphone Audio Exposure. Headphone Audio Exposure measures audio exposure from PLDs in units of dB(A). It is supported across various devices: iPhone from iOS 13.0+, iPad from iPadOS 13.0+, and Apple Watch from watchOS 6.0+. As new records are generated when users initiate PLD sessions or when audio exposure levels change during continuous use, data is sampled irregularly. Extracted information includes start/end timestamps, audio exposure level (dB(A)), device name, manufacturer, and data source, with historical data availability enabled by using the same phone for a long time or by iCloud storage persistence across device upgrades.

The survey responses from non-iPhone users were stored in the Sosci Survey portal and exported at the end for analysis. For iPhone users, survey responses and historical data were transmitted from the app to a secure server via an API endpoint implemented in a Flask-based backend. All uploaded data were stored in an AES-encrypted format to ensure participant confidentiality. No personally identifiable information was collected, and stored file names were based on the \emph{identifierForVendor} key from Apple and the timestamp. The \emph{identifierForVendor} key is an alphanumeric string that remains constant for apps from the same vendor on the same device, as long as the app remains installed; however, reinstallation yields a different value.





\section{Dataset}
\label{sec:dataset}


Over a period of 8 months, a total of \participants{} participants completed the study questionnaire (\male{} male, \female{} female, \diverse{} diverse; age range: \ageMin{}--\ageMax{} years, $M = \ageMean{}$ years, $SD = \ageStd{}$). The participants were predominantly based in Germany ($\germanGroup{}\,\%$), with the remaining participants coming from 14 other countries.

\subsection{Data Cleaning and Preprocessing}
\label{subsec:preprocessing}

As outlined earlier, non-iPhone users ($n = \androidParticipants$) completed the survey-only participation through the Sosci Survey platform, while the iPhone participants ($n = \totalIphone$) completed both the survey and, additionally, had the option to donate historical health data through our custom app. Two health data files were corrupted and therefore excluded from analysis. Among the participants with iPhone, thirty-three participants did not share their health data. Headphone Audio Exposure data were available for \iPhoneParticipants{} participants, who together recorded a total of \totalSessions{} sessions during 2020 to 2024 with a median duration of \medianDuration{} minutes, with an interquartile range of \qOneDuration{} to \qThreeDuration{} minutes. The gap between the number of people donating the data and the actual number of participants with data is due to either a lack of data within the study period or device version-based constraints.

Due to high variability in the data across participants, data cleaning and preprocessing were necessary. Since Headphone Audio Exposure data are irregularly sampled and device-dependent, we applied standardized preprocessing to harmonize temporal granularity, reduce noise, and ensure comparability across participants. The study period spanned 2020--2024, based on data availability; changes to this timeline are noted where relevant. Since data collection took place in 2025, the analysis was restricted to the period 2020–2024 to avoid data attrition near the time of participation.

\subsubsection{Data Preprocessing}
HealthKit API provides data in UTC timestamps; however, using them directly would obscure granular temporal trends in user behavior. All UTC timestamps were converted to participants' local time zones based on their reported country of residence, as indicated in the survey responses. To address fragmentation from volume adjustments, we merged entries separated by less than 120 seconds and computed time-weighted average volumes for the merged periods. Sessions spanning midnight were split across day boundaries. Finally, the absence of data for a given time span was assumed to indicate no usage. Apart from these adjustments, no implausible values were detected. Importantly, historical data analyses describe the behavior of participants who shared iPhone PLD records, rather than the whole PLD user population.

Due to the irregular sampling across the metric, the overall distributions are skewed. Therefore, group comparisons used non-parametric tests, while repeated-measures analyses used generalized estimating equations with cluster-robust standard errors, which remain consistent under non-normal residuals. The standard significance level ($\alpha = \mainAlpha{}$) was applied, with family-wise Bonferroni corrections for multiple testing at each occasion. Effect sizes for Wilcoxon tests were computed as $r_{W}$ and interpreted according to Cohen's effect size convention \cite{cohen2013statistical}. For correlation analyses, Spearman's correlation ($\rho$) was used when at least one ordinal variable was involved and were interpreted using Cohen's correlation convention \cite{cohen2013statistical}.

\section{Data Analyses}
\label{sec:data_analyses}
This section outlines the procedures and presents the findings of all analyses conducted to address the research questions.

\subsection{RQ1: Patterns of PLD Use and ANC Adoption}
\hypertarget{RQ_I}{Understanding} the contexts in which PLDs are used is crucial for developing user-specific applications. Therefore, we analyzed survey responses regarding PLD and ANC usage to characterize their practical applications. As all participants completed the survey, responses from the entire sample (iPhone and non-iPhone users;  $N = \participants{}$) were included in the analysis.

\begin{figure}[t!]
\centering
\includegraphics[width=\linewidth]{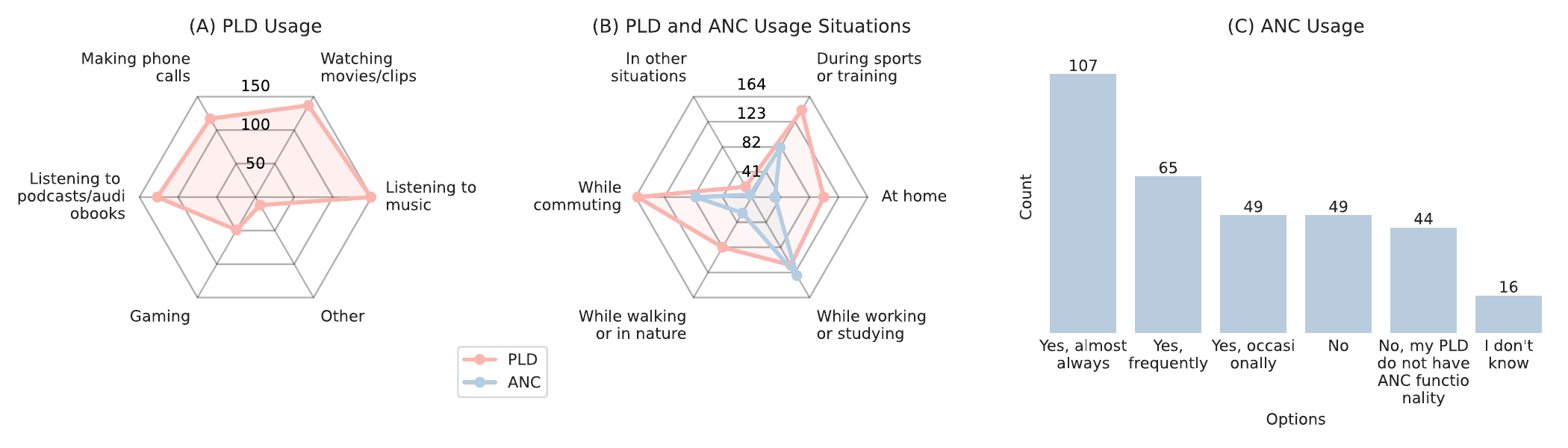}
\caption{Everyday contexts of PLD and ANC usage ($N = \participants{}$). (A) Activities performed while using PLDs; (B) Contexts in which participants typically use PLDs and ANC; (C) ANC adoption. (A) and (B) show the number of responses for each option in a multiple-selection question.}
\Description
[Survey results on PLD and ANC usage (purposes, contexts, ANC adoption)]{
The figure has three panels. Panel A (PLD Usage) shows a radar plot of the number of participants reporting PLD use across six activities. The highest reported use is for listening to music, followed by watching movies or clips and making phone calls. Lower but notable usage is seen for gaming. The “other” category has the least reported use. The values radiate outward, with larger distances from the center indicating more frequent use. Panel B (PLD and ANC Usage Situations) shows a radar plot comparing two lines: a red line for PLD usage and a blue line for ANC usage, across six everyday situations. Both PLD and ANC usage peak during sports or training and while working or studying. Overall, the red line (PLD) extends farther than the blue line (ANC) in most situations, indicating that PLD use is more common overall, though ANC follows similar situational patterns. Panel C shows a bar chart of ANC adoption. The highest group is "Yes, almost always" (107 participants), followed by "Yes, frequently" (65), "Yes, occasionally" (49), "No" (49), "No, my PLD does not have ANC functionality" (44), and "I don’t know" (16).
}
\label{fig:survey_res}
\end{figure}

To investigate everyday PLD usage, we analyzed the purposes for which participants typically use PLDs (\hyperlink{q5}{Question 5}). The results indicate that PLDs were most commonly used to listen to music (\pldMusic{}\,\%), watch movies (\pldMovie{}\,\%), listen to podcasts (\pldPodcast{}\,\%), and make calls (\pldCall{}\,\%). Gaming (\pldGaming{}\,\%) was less frequently chosen, while an additional \pldOthers{}\,\% selected the ``Other'' option, most often sleeping and noise suppression. These results are visualized in \autoref{fig:survey_res}.

In addition, many modern PLDs feature ANC, which suppresses environmental noise, reduces sound exposure, and supports hearing health~\cite{hodgetts2007effects, seol2022influence}. ANC fundamentally changes how loud and where people use PLDs; however, little is known about its adoption. Assessing ANC adoption among participants (\hyperlink{q8}{Question 8}) revealed that more than half (\ancActive{}\,\%) reported actively using ANC, either almost always or frequently. In contrast, \ancDidNot{}\,\% did not use ANC, with \ancNoSupport{}\,\% noting that their PLDs do not support the feature. Finally, \ancNotAware{}\,\% reported being unaware of ANC. 

Further, we explored the contexts in which participants reported wearing PLDs (\hyperlink{q6}{Question 6}) and, for ANC users, the situations in which ANC was enabled (\hyperlink{q9}{Question 9}). PLDs were most commonly used while commuting (\pldCommuting{}\,\%), during sports (\pldSport{}\,\%), for work or study (\pldWork{}\,\%), at home (\pldHome{}\,\%) and while walking or nature (\pldWalkNature{}\,\%). Other PLD use situations included falling asleep and shopping. For ANC, out of the \totalANCUseParticipants{} participants who use them, the results revealed a similar but more selective pattern: ANC was especially common during work/study (\ancWork{}\,\%), during sports (\ancSport{}\,\%), and while commuting (\ancCommuting{}\,\%).

\subsection{RQ2: Temporal Patterns of PLD Usage}
\hypertarget{RQ_II}{Understanding} temporal patterns of PLD usage provides realistic expectations for earable application designs. To understand temporal variations in PLD usage behavior, we conducted comprehensive analyses across multiple time scales, from within-week variation to long-term trends spanning five years. This multi-scale temporal analysis enabled us to identify patterns at different granularities and to understand how usage behavior varies across natural time boundaries, such as weeks and years. This analysis used historical data from iPhone users ($n=\iPhoneParticipants{}$). The median participant logged no listening at all on \zeroDayRate{}\,\% of days in 2024 (IQR 23.09–73.91\,\%), with a median of 1.52 listening episodes across all days.

\subsubsection{PLD Usage Trend Across Week}
To evaluate PLD usage patterns across the week, we aggregated average usage duration by weekday. Next, the day-wise data was further grouped into weekdays and weekends (Saturday and Sunday) and then compared. To examine whether PLD usage differed between weekdays and weekends, we used a multivariate GEE model with a Gaussian family, identity link, and an independence working correlation structure that analyzed two outcomes simultaneously: mean daily usage duration (total hours) and mean daily session count. Both outcomes were z-standardized before analysis so they could be directly compared on a common scale. We used an independence working correlation structure in the GEE model to minimize assumptions about temporal dependence. The aggregated outcome was right-skewed, so we based inference on cluster-robust standard errors, which remain consistent under non-normal residuals and a misspecified working correlation. The data were averaged so that each participant had one observation per day type per outcome, resulting in a total of \weekendObservations{} observations from \iPhoneParticipants{} participants.

The interaction between day type and outcome type was significant ($b = \jointB{}$, $z = \jointZ{}$, $p < \jointp{}$), indicating that the weekend effect manifested differently across the two outcomes. Overall combined engagement was significantly lower on weekends than on weekdays ($b = \sessionB{}$, $z = \sessionZ{}$, $p = \sessionp{}$), yet this reduction was less pronounced for total usage duration than for session count. On weekdays, the two outcomes did not differ significantly from each other ($b = \durationB{}$, $z = \durationZ{}$, $p = \durationP{}$), confirming a comparable baseline. Together, these results suggest a compensatory pattern: on weekends, participants initiated fewer listening sessions but listened for longer when they did, leaving cumulative daily exposure largely unchanged.

\subsubsection{PLD Usage Trend Across Years}
We further analyzed long-term trends to assess PLD adoption over the years. Since the data availability across years varied widely among participants, the total PLD usage of participants with less than 5 hours in a year was set to null to remove differences due to data availability for a given year. A GEE model with a Gaussian family, identity link, and an autoregressive working correlation structure was used to account for the temporal dependence observed in the longitudinal data, in which usage in adjacent years was more strongly correlated than in distant years. The aggregated outcome was right-skewed, so we based inference on cluster-robust standard errors, which remain consistent under non-normal residuals and a misspecified working correlation. This analysis included \yearTotalParticipants{} participants, as 2 participants had < 5 hours of data in all years. The model revealed a significant increase in mean daily usage over the study period, based on \yearTotalObservations{} participant-year observations. Compared to 2020 ($M = \yearTwentyM{}$ min), daily PLD usage increased significantly in 2022 ($M = \yearTwentyTwoM{}$ min; $b = \yearTwentyTwob{}$, $z = \yearTwentyTwoz{}$, $p = \yearTwentyTwop{}$), 2023 ($M = \yearTwentyThreeM{}$ min; $b = \yearTwentyThreeb{}$, $z = \yearTwentyThreez{}$, $p = \yearTwentyThreep{}$), and 2024 ($M = \yearTwentyFourM{}$ min; $b = \yearTwentyFourb{}$, $z = \yearTwentyFourz{}$, $p = \yearTwentyFourp{}$). In 2021 ($M = \yearTwentyOneM{}$ min; $b = \yearTwentyOneb{}$, $z = \yearTwentyOnez{}$, $p = \yearTwentyOnep{}$), there was a small rise in usage trend compared to 2020, but this result was not significant. Overall, PLD listening increased sharply after 2020, with a slight dip through 2023, before rising again in 2024.

\subsection{RQ3: Psychological Factors Related to PLD Usage Patterns}
\hypertarget{RQ_III}{Understanding} the psychological factors related to PLD usage is crucial for both explaining variability in usage and recruiting participants for in-the-wild studies. To this end, we analyzed the survey responses from both iPhone and non-iPhone users ($N = \participants{}$).

We examined how sensation seeking tendencies, personality traits, and perceptions of using PLD in social situations relate to usage patterns. The BSSS and BFI-10 scores were first computed from 5-point Likert-scale responses, yielding continuous scores ranging from 1 to 5 (1 = lowest). These scores were then correlated with the survey questions on PLD usage frequency (\hyperlink{q4}{Question 4}), usage duration (\hyperlink{q7}{Question 7}), and music listening duration (\hyperlink{q10}{Question 10}). Spearman's rank correlation ($\rho$) was computed to compare the ordinal survey responses with the questionnaire scores, and the results were interpreted according to Cohen's convention for interpreting correlations \cite{cohen2013statistical}. Furthermore, attitudes towards using PLD in social settings (\hyperlink{q11}{Question 11}), perceptions of social isolation (\hyperlink{q12}{Question 12}), and psychological shielding (\hyperlink{q13}{Question 13}) while using PLD were also evaluated to get a better picture of PLD usage associated with psychosocial factors.

\begin{figure}[t!]
\centering
\includegraphics[width=\linewidth]{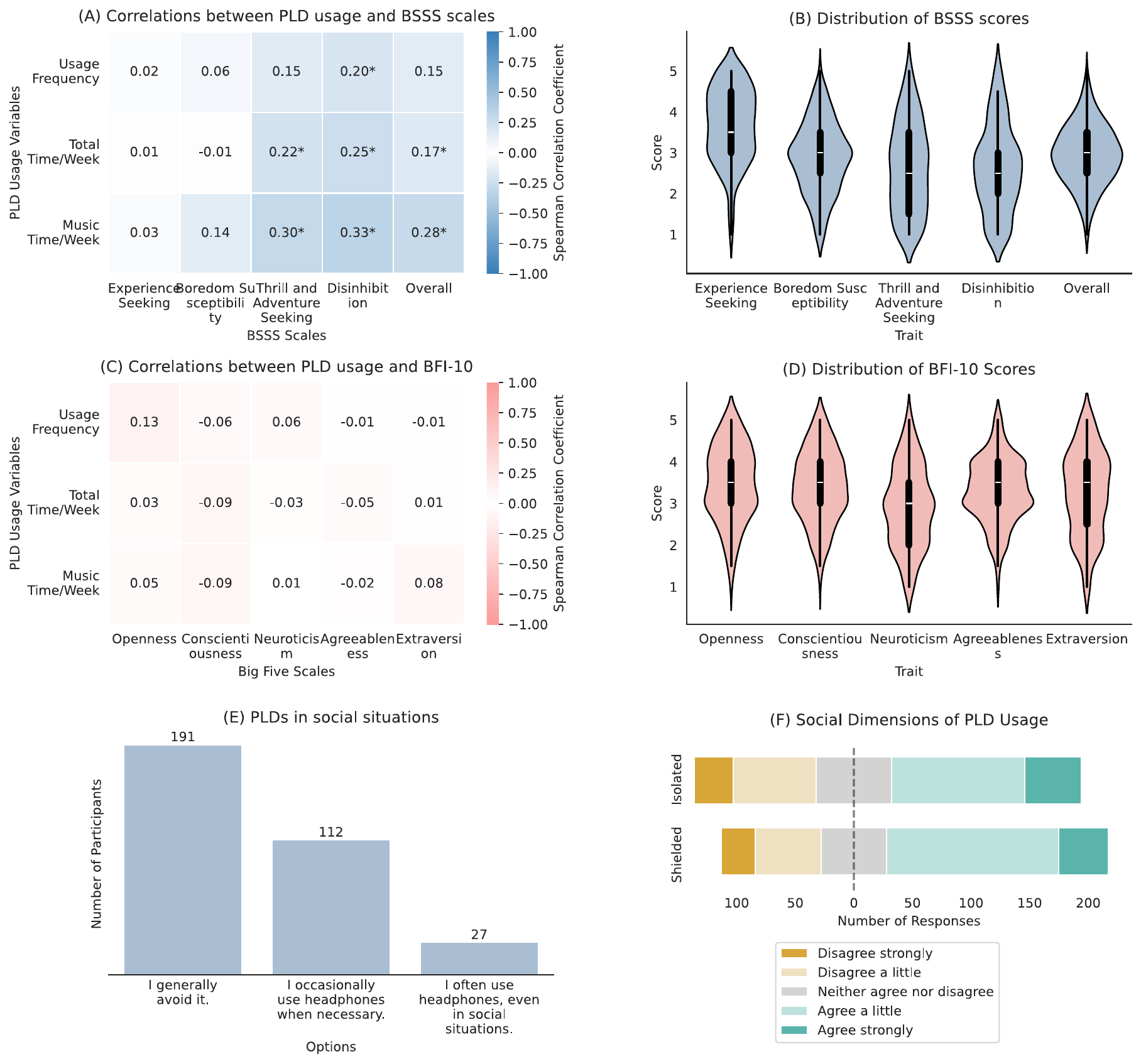}
\caption{Influence of psychological factors on PLD usage ($N = \participants{}$). 
(A) Spearman correlations between BSSS facet scores  and PLD usage variables;
(B) Distribution of BSSS scores; 
(C) Spearman correlations between BFI-10 trait scores and PLD usage variables; 
(D) Distribution of BFI-10 scores; 
(E) Participants' self-reported attitudes toward using PLDs in social situations; 
(F) Social perceptions of PLD use: feelings of isolation and psychological shielding. The asterisk (*) sign in (A) and (C) represents significant correlations with $\alpha = \psyBonferroni{}$.}
\Description
[Influence of psychological factors on PLD usage.]{The figure presents six panels. Panel A shows a heatmap of Spearman correlations between BSSS facet scores (experience seeking, boredom susceptibility, thrill and adventure seeking, Disinhibition, and overall) and PLD usage variables (usage frequency, total time per week, music time per week). Correlations are medium, but a significant positive relationship is observed between Disinhibition and both PLD usage frequency, duration, and music listening duration. Small to medium correlations are observed between music listening and overall BSSS score and thrill and adventure seeking. Further, total usage time has a small to medium correlation with overall BSSS score and thrill and adventure seeking. Panel B illustrates the distributions of BSSS scores using violin plots. All traits cluster around the scale midpoint of 2.5, with balanced spreads, indicating moderate sensation seeking levels among participants, with experience seeking slightly skewed. Panel C shows a heatmap of correlations between BFI-10 personality traits (openness, conscientiousness, neuroticism, agreeableness, and extraversion) and PLD usage variables. These correlations are generally weak, and none are statistically significant. Panel D depicts the distributions of BFI-10 scores using violin plots. These are also centered around the midpoint, reflecting balanced personality profiles across the sample. Panel E shows a bar chart of participants’ attitudes toward PLD use in social situations: 191 participants reported they generally avoid it, 112 occasionally use PLDs when necessary, and 27 often use PLDs even in social contexts. Panel F presents social perceptions of PLD use in a diverging bar chart. Responses are spread across agreement and disagreement, with some participants reporting feelings of social isolation when using PLDs, while others report a sense of psychological shielding.}
\label{fig:psy}
\end{figure}

\subsubsection{Sensation Seeking Tendencies and PLD Usage} The correlation results between PLD usage and BSSS scores are shown in \autoref{fig:psy} (A), and the overall distribution of these scores is presented in \autoref{fig:psy} (B). The significance level was adjusted to $\alpha < \psyBonferroni{}$ using Bonferroni correction because multiple correlation comparisons were performed between the scores and the PLD usage variables. Experience Seeking ($Md = \medExpSeek{}$) showed the highest median scores, while medians of Thrill and Adventure Seeking ($Md = \medThrill{}$) and Disinhibition ($Md = \medDisinhibition{}$) were precisely at the midpoint with greater variability. Both Boredom Susceptibility ($Md = \medBoredom{}$) and the overall BSSS score ($Md = \medOverall{}$) were slightly above the midpoint, indicating generally elevated sensation seeking tendencies. The overall sensation seeking scores showed a small to medium correlation with music listening time ($\rho = \overallMusicTime{}, p < \overallMusicTimep{}$) and weekly usage time ($\rho = \overallPLDUsage{}, p = \overallPLDUsagep{}$). Disinhibition demonstrated a small to medium correlation with usage frequency ($\rho = \disinhibitionPLDFreq{}, p < \disinhibitionPLDFreqp{}$) and weekly usage time ($\rho = \disinhibitionPLDUsage{}, p < \disinhibitionPLDUsagep{}$), but a medium correlation for music listening time ($\rho = \disinhibitionMusic{}, p < \disinhibitionMusicp{}$). Thrill and adventure seeking showed a small to medium correlation with weekly usage time ($\rho = \thrillPLDUsage{}, p < \thrillPLDUsagep{}$) and a medium correlation with music time ($\rho = \thrillMusic{}, p < \thrillMusicp{}$). These patterns suggest that individuals higher in sensation seeking, particularly those prone to Disinhibition and Thrill and Adventure Seeking, are associated with pronounced PLD usage.

\subsubsection{Personality Traits and PLD Usage} The results for the Big Five personality traits are shown in \autoref{fig:psy} (C), with score distributions in \autoref{fig:psy} (D). The significance level was adjusted to $\alpha < \psyBonferroni{}$ using Bonferroni correction because multiple correlation comparisons were performed between the scores and the PLD usage variables. Median values for Openness ($Md = \medOpenness{}$), Conscientiousness ($Md = \medConscientiousness{}$), Agreeableness ($Md = \medAgreeableness{}$), and Extraversion ($Md = \medExtraversion{}$) were above the midpoint of the 1–5 scale, with relatively symmetric distributions around those central values. Neuroticism ($Md = \medNeuroticism{}$) displayed a lower median, closer to the midpoint of the scale, and a wider spread compared to the other traits. Overall, the score distributions approximate a normal distribution with modest variation across traits. BFI-10 personality traits showed no significant associations with usage patterns.

\subsubsection{Perceptions and Overall Attitudes Toward Using PLDs in Social Situations}
\label{subsubsec:social}
Understanding the social factors related to PLD usage is crucial for designing earable applications with practical relevance. While personality characteristics offer one lens, PLD use is also linked to social and emotional contexts. The question on social isolation assessed whether participants perceived PLDs as creating barriers to social interaction, while the question on shielding examined whether participants viewed PLDs as providing psychological protection from their environment. Participants' responses were scored on a 5-point Likert scale. Alongside the overall percentages of agreement and disagreement, response polarization (ratio of “strongly agree” and “strongly disagree” responses to total responses) was reported to convey the strength of respondents’ attitudes. Results show that most participants reported social awareness when using PLDs (\autoref{fig:psy} (E)). A majority (\totalSocialAvoidance{}\,\%) either avoided using PLDs in social situations ($\completeAvoidance{}\,\%$) or used them only when necessary ($\mostlyAvoidance{}\,\%$). 

To further understand PLD usage in social situations, we analyzed the responses on feeling social isolation or shielding while using PLDs (\autoref{fig:psy} (F)), about half (\agreedIncreasedIsolation{}\,\%) agreed that PLDs increased feelings of social isolation, while a third (\disagreedIncreasedIsolation{}\,\%) disagreed. The overall polarization for social isolation is \polarizationIsolation{}\,\%, suggesting that most participants did not express strong opinions, with relatively few selecting extreme agreement or disagreement. 
Conversely, nearly half (\agreedIncreasedShielding{}\,\%) reported that PLDs provided social shielding, compared to \disagreedIncreasedShielding{}\,\% who disagreed. The overall polarization for social shielding is \polarizationShielding{}\,\%, again indicating low polarization, with responses concentrated around moderate agreement/disagreement. Follow-up analyses examined the relationship between feelings of isolation and shielding, with significance level adjusted to $\alpha < \psySocialBonferroni{}$ using Bonferroni correction because of multiple correlation comparisons. The two constructs have a large positive correlation ($\rho$ = \shieldVsIsolation{}, $p < \shieldVsIsolationp{}$ ), indicating that participants who felt isolated while wearing PLDs also tended to feel shielded. This pattern was reflected in the crosstabulation: extreme mismatches between the two scales were rare, suggesting substantial co-occurrence (See \autoref{sec:agreement}). To better understand how feelings of social isolation and shielding relate to PLD use in social settings, a correlation analysis was performed. Responses of feeling social isolation while using PLD showed a small to medium negative correlation ($\rho$ = \isolationVsSocial{}, $p = \isolationVsSocialp{}$), indicating that higher perceived isolation was associated with lower reported PLD use in social settings. However, feelings of shielding did not show any significant correlation to PLD usage in social settings ($\rho = \shieldVsSocial{}$, $p = \shieldVsSocialp{}$).
\subsection{RQ4: Demographics and PLD Usage}
\hypertarget{RQ_IV}{Understanding} the demographic factors related to PLD usage is important for recruiting participants for in-the-wild studies and designing user-specific studies. To study this, we again considered both survey and historical data; therefore, we included only participants with both data ($n = \iPhoneParticipants{}$).

To investigate potential demographic influences on PLD usage duration (average daily hours) and volume level (average dB(A)), we conducted comparative analyses across gender and age groups. Demographic information, including age (\hyperlink{q2}{Question 2}) and gender (\hyperlink{q3}{Question 3}), was collected through surveys. For single-group comparisons, the Mann-Whitney $U$ test was used to test the hypothesis, while for multi-group comparisons, the Kruskal-Wallis $H$ test was followed by post-hoc pairwise Mann-Whitney $U$ tests with significance levels corrected after Bonferroni.

\begin{figure}[!t]
\centering
\includegraphics[width=\linewidth]{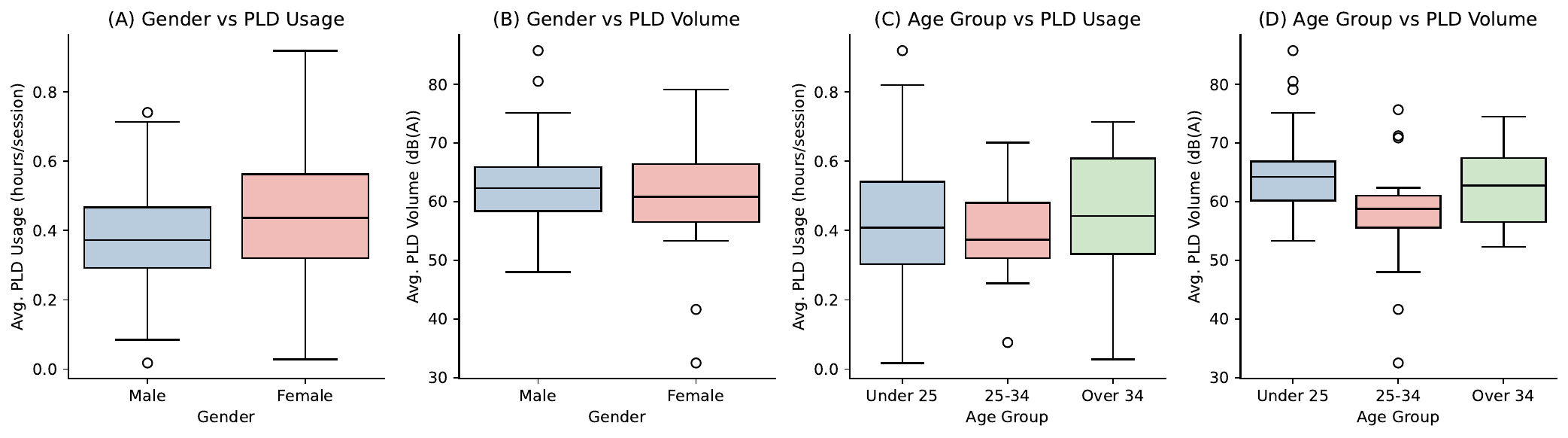}
\caption{Impact of demographics on PLD usage and volume ($n = \iPhoneParticipants{}$).
(A) Average PLD usage duration across gender groups;
(B) Average PLD volume across genders;
(C) Average PLD usage duration across age groups;
(D) Average PLD volume across age groups.}
\Description
[Impact of demographics on PLD usage and volume.]{The figure contains four boxplots. Panel A compares average PLD usage duration (in hours) across gender. Both males and females show similar median usage, around 0.37–0.42 hours, though females show slightly higher variability, with some sessions extending longer. Panel B compares average PLD listening volume (in dB(A)) across genders. Males and females have very similar median volumes, slightly above 60 dB(A), with little difference overall. Panel C compares average PLD usage duration across three age groups: under 25, 25–34, and over 34. All three groups have similar average listening duration around 0.4 hours. Panel D compares average listening volume across the same age groups. The under-25 group shows a higher median listening volume, around 65 dB(A), compared to the 25–34 group at about 59 dB(A) and the over-34 group at about 62 dB(A). Outliers are present in several groups. 
}
\label{fig:demographics_usage}
\end{figure}

\subsubsection{Gender and PLD Usage}
To compare PLD usage across gender groups, only the "male" ($n = \maleNumber{}$) and "female" ($n = \femaleNumber{}$) groups were considered, as there were no donations from the "diverse" group. Male and female participants demonstrated similar PLD usage across both measures examined.  PLD usage duration (\autoref{fig:demographics_usage} (A)) did not reveal a significant difference between female ($Md = \femaleMedUsageTime{}$ min) and male ($Md =  \maleMedUsageTime{}$ min; $U = \genderUsageTimeU{}, p = \genderUsageTimeP{}$, $r_{W} = \genderUsageTimeR{}$) participants. Similarly, volume levels (\autoref{fig:demographics_usage} (B)) did not reveal a significant difference between female ($Md = \femaleMedUsageVolume{}$ dB(A)) and male ($Md = \maleMedUsageVolume{}$ dB(A); $U = \genderUsageVolumeU{}, p = \genderUsageVolumeP{}, r_{W} = \genderUsageVolumeR{}$).

\subsubsection{Age and PLD Usage}
Age data were grouped into "Under 25" (18-24; $n = \UnderCount{}$), "25-34" ($n = \MidCount{}$), and "Over 34" (> 34; $n = \OldCount{}$) because the study only included adults aged 18 and above, and participants older than 34 were relatively few, so combining them into a single "Over 34" category ensured adequate sample sizes for analysis. The significance level for post hoc tests was set to 0.02 after Bonferroni correction for multiple comparisons across the three age groups. Results show that PLD usage duration (\autoref{fig:demographics_usage} (C)) showed no significant difference between the groups ($H = \ageUsageTimeH{}, p = \ageUsageTimeP{}$). In contrast, volume levels (\autoref{fig:demographics_usage} (D)) showed significant age-related differences across the three groups ($H = \ageVolumeH{}, p = \ageVolumeP{}$). Participants under 25 years ($Md = \UnderMedVolume{}$ dB(A)) listened at significantly higher volumes than the 25-34 age group ($Md = \MidMedVolume{}$ dB(A)) with small to medium effect ($U = \ageVolumeUnderMidU{}, p = \ageVolumeUnderMidP{}, r_{W} = \ageVolumeUnderMidR{}$). However, participants over 34 did not show a significant difference between the groups and exhibited intermediate volume levels ($Md = \OldMedVolume{}$ dB(A)).

\subsection{RQ5: PLD Listening Levels}
\hypertarget{RQ_V}{Safety} is a key concern for PLD use. To accurately assess potential risks, it is essential to understand realistic exposure patterns. This section presents the findings from our study, which examined PLD usage patterns among participants who donated objective behavioral data ($n = \iPhoneParticipants{}$). Overall, the median volume was \medianVolume{} dB(A), with an interquartile range of \qOneVolume{} to \qThreeVolume{} dB(A).

We conducted a comprehensive assessment of participants' listening volume and their adherence to the WHO guidelines. Exposure was calculated using the formula $40 * 2^{-(L-80)/3}$ hours per week, where L is the listening level in dB(A). This formula derives from the WHO safe listening limit of 40 hours per week at volume less than 80 dB(A), combined with a 3 dB exchange rate under which the allowable exposure time halves for every 3 dB increase in level \cite{WHO2019SafeListening}. Weekly dose contributions exceeding 100\% indicated guideline non-compliance. The accuracy of the Apple Health calibration is based on the sound level delivered to the ear canal in dB(A). While this is well calibrated for Apple and Beats PLDs, it is less precise for third-party or wired devices. For third-party devices, Apple Health stores the device volumes instead.

\begin{figure}[!t]
\centering
\includegraphics[width=\linewidth]{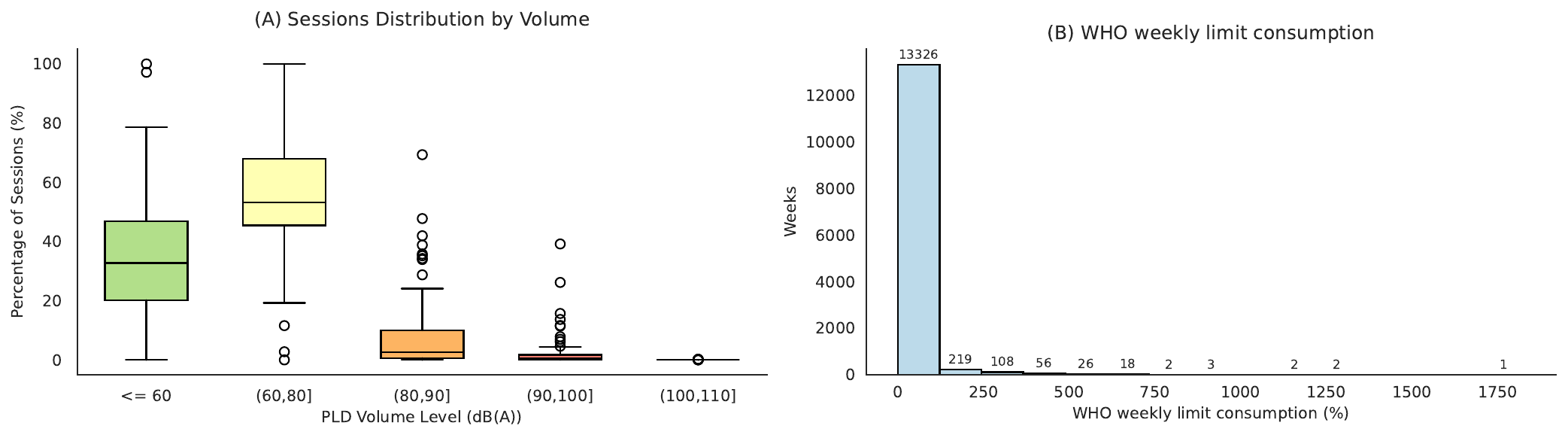}
\caption{PLD listening volume trends ($n = \iPhoneParticipants{}$). (A) Distribution of listening sessions by PLD volume levels, showing percentages across different dB(A) ranges; (B) Distribution of WHO weekly limit consumption percentage (participant-week = 13734).}
\label{fig:WHO}
\Description[WHO safe listening compliance and risk categorization of PLD usage.]{The figure presents two panels. Panel A illustrates the distribution of listening sessions by PLD volume levels. The majority of sessions fall below 60 dB(A) and 61-80 dB(A), smaller proportions fall between 81–90 dB(A), and only a very small percentage extend into higher categories (91–100 dB(A) and 100–110 dB(A)). Panel B shows the distribution of the WHO weekly limit consumption (\%). Most participant-weeks fall within the recommended limits.}
\end{figure}

\subsubsection{Volume Trend Across Years}
We further examined long-term trends in listening intensity. As mentioned earlier, the data availability across years varied widely among participants; hence, the total PLD usage of participants with less than 5 hours in a year was set to null to remove differences due to data availability for a given year. A Gaussian GEE with an autoregressive working correlation structure was used to analyze the change in mean weekly volume from 2020 to 2024. The aggregated outcome was left-skewed, so we based inference on cluster-robust standard errors, which remain consistent under non-normal residuals and a misspecified working correlation. The analysis, comprising \volumeObservations{} observations from \totalParticipantsVolume{} participants, revealed a significant initial decline in listening volume following the baseline year. Compared to 2020 ($M = \volumeTwenty{}$ dB(A)), average weekly volume decreased significantly in 2022 ($M = \volumeTwentyTwo{}$ dB(A); $b = \volumeTwentyTwob{}$, $z = \volumeTwentyTwoZ{}$, $p = \volumeTwentyTwop{}$). While volume remained lower in 2021 ($M = \volumeTwentyOne{}$ dB(A); $b = \volumeTwentyOneb{}$, $z = \volumeTwentyOneZ{}$, $p = \volumeTwentyOnep{}$), 2023 ($M = \volumeTwentyThree{}$ dB(A); $b = \volumeTwentyThreeb{}$, $z = \volumeTwentyThreeZ{}$, $p = \volumeTwentyThreep{}$) and 2024 ($M = \volumeTwentyFour{}$ dB(A); $b = \volumeTwentyFourb{}$, $z = \volumeTwentyFourZ{}$, $p = \volumeTwentyFourp{}$) relative to the baseline, these differences did not reach statistical significance. Overall, listening intensity showed a stabilizing or slightly downward trend after 2022.

\subsubsection{Listening Sessions Distributed Across Sound Exposure Categories} Further, we analyzed proportional exposure across sound level bands ($\leq$ 60, (60,80], (80,90], (90,100], (100,110] dB(A)). For each participant, we calculated the proportion of sessions in each band to assess safe versus unsafe listening, revealing exposure patterns and level preferences associated with compliance or non-compliance with the WHO recommendations (\autoref{fig:WHO} (A)). Most of the PLD listening occurred $\leq$ 80 dB(A). Exposure to (80,90] dB(A) ($Md = \medEightyOneToNinety{}\%$) and (90,100] dB(A) ($Md = \medNinetyOneToHundred{}\%$) were less frequent. Listening above 100 dB(A) ($Md = \medOverHundred{}\%$) was rare: only \partOverHundred{}\,\% of the participants had any exposure in the (100,110] dB(A) range and no sessions with volume above 110 dB(A). Similarly, only \partOverNinety{} participants listened to more than 5\,\% of their sessions at volume >90 dB(A).

\subsubsection{WHO Safe Listening Recommendation Adherence}
To determine whether the WHO safe listening recommendation is followed, we evaluated the percentage of weekly exposure below the WHO-recommended limit. Only \participantWeekExceed{}\,\% of the participant-weeks have exceeded the WHO recommendation (\autoref{fig:WHO} (B)), with \participantAtleastOneWeekExceed{} participants having at least 1 week of exceeding the recommended exposure.

\section{Discussion}
Our study characterizes everyday PLD use by combining self-reports from \participants{} adults with multi-year, passively logged audio-exposure records from \iPhoneParticipants{} iPhone users. Because the design is observational, every relationship we report is correlational; we interpret them as associations and patterns of co-occurrence, not as effects.

\subsubsection{Contextual and acoustic control as a design surface}
Prior work established that listeners raise playback levels in noisy environments \cite{hodgetts2007effects} and expressed strong demand for better noise control in everyday PLDs, while studies of ANC characterized its value in specific settings such as open-plan offices and clinical care \cite{bhatia2025revolutionizing, mueller2022using}. Our survey responses indicate that users employ PLD and ANC as active tools for managing acoustic boundaries rather than merely as passive listening or generic noise reduction mechanisms. ANC is most frequently enabled in contexts such as commuting, work, and sport or training, where environmental noise or cognitive demands motivate stronger control over auditory input. In contrast, ANC is used less often when walking in nature or relaxing at home, suggesting a deliberate weighing of environmental control against situational awareness, enjoyment of ambient sound, or social engagement. These nuanced adoption patterns highlight that ANC is part of a broader strategy for modulating comfort, focus, and privacy across everyday situations, and they raise important design questions for earables: how can ear-based devices adapt acoustic environments to shifting activities and preferences while maintaining transparency about sensory boundaries for both psychological and practical benefits? At the same time, our self-report data indicate that ANC remains underutilized, implying that awareness-building and safe-listening guidance around ANC could play a larger role in mitigating high sound levels \cite{hodgetts2007effects,seol2022influence}.

\subsubsection{Wear time is structured, not continuous}
As mentioned earlier, earable systems for continuous mobile-health sensing and context-aware inference generally presuppose stable, extended wear. Though \citet{neitzel2022toward} shared preliminary results of passively logged PLD exposure at large scale, their focus was on sound exposure. We extend that passive-data approach along the temporal dimension that it left unaddressed. Almost 48\,\% of the participant-days did not have any usage in 2024. Further, combined daily engagement was lower on weekends than on weekdays, with the reduction less pronounced for total listening duration than for the number of sessions, and daily usage was higher in the later years of the study window.

Earable applications that presuppose all-day wear are unlikely to obtain the coverage they assume. The more defensible design target is intermittent wear as the default: systems should model naturally occurring usage windows — commuting, focused work, exercise — rather than a flat baseline, and in-the-wild earable studies should plan data collection around weekly rhythms instead of assuming uniform daily availability. Treating intermittent coverage as the norm, through event-triggered sensing and imputation that is explicit about its gaps, is likely to be more robust than engineering for uninterrupted wear.

\subsubsection{Individual differences in sustained engagement}
Earlier works associated PLD use with anxiety and arousal regulation \cite{ekcsi2019headphone} and reported that people higher in sensation seeking show greater physiological arousal to PLD listening \cite{kallinen2007comparing}. Our work extends this by looking deeper into the facets of sensation seeking. Participants with high disinhibition or thrill-and-adventure seeking scales showed small to medium positive correlations with self-reported weekly usage and medium positive correlations with self-reported music listening.

Applications that depend on prolonged, consistent wear should anticipate heterogeneous adherence rather than a single expected wear profile. The association between sensation seeking facets and heavier reported use suggests that trait information, where users consent to provide it, could help set realistic per-user coverage expectations and inform recruitment for longitudinal deployments, while users with lower baseline engagement may be better served by shorter monitoring windows. Because these relationships are correlational and small to medium, we frame trait-aware personalization as a hypothesis for earable designers to test, not a targeting rule to apply.

\subsubsection{Social contexts as systematic coverage gaps}
Prior work framed PLDs as tools people use to withdraw from social contact and structure their routines \cite{harshitha2017survey}. A large majority of participants reported avoiding PLD use in social situations or limiting it to specific needs; perceived isolation and perceived shielding co-occurred; and feeling isolated while using PLDs has a small negative association with using them in social settings. 

If users routinely remove ear-worn devices during interpersonal encounters, applications that assume continuous social coverage — conversational well-being analysis \cite{min2018cross}, or monitoring around socially salient moments — face predictable, non-random wear gaps precisely when the events of interest occur. Designers face a choice between targeting solitary or routine contexts where continuous wear is naturalistic and developing intermittent-sensing approaches that tolerate social gaps. The tendency to remove devices in social settings should be treated as a major design constraint, and socially aware interaction techniques should respect users' evident wish to modulate availability.

\subsubsection{Age relates to PLD volume while gender did not}
Prior work documented a genuine disagreement: \citet{torre2008young} report that men listen at higher levels and for longer, \citet{fligor2014cultural} find no gender differences. In the logged data, we observed no significant gender differences in logged listening duration or volume, which cautions that the differences reported in some self-report and laboratory work may not persist in naturalistic use. For age, younger participants (under 25) listened at higher median volumes than the 25–34 group with no accompanying difference in duration, consistent with the hearing-health literature linking younger listeners to higher recreational exposure \cite{jokitulppo1997estimated}.

The higher volumes among younger listeners point to age-aware safe-listening features — adaptive loudness defaults calibrated to higher-risk groups — as a concrete design direction, provided such features rely on each user's measured exposure rather than group membership alone.

\subsubsection{In-the-wild exposure and the case for earable safe-listening}
Clinical work has tied high recreational exposure to tinnitus and early hearing loss \cite{clark1992hearing,jokitulppo1997estimated,ramage2019tinnitus}, and survey studies reported substantial proportions of users listening at high or maximum volume but, as we noted, almost all quantify exposure from self-reported volume. \citet{neitzel2022toward} shared that the mean PLD volume among 121,010 participants was around 62.7 dB(A) which aligns with our analysis (M: 63.88 dB(A), Md: 64.84 dB(A)). Most logged listening occurred at or below 80 dB(A), and a small proportion of participant-weeks (\participantWeekExceed{}\,\%) exceeded the WHO weekly recommendation, concentrated in a subset of participants. The \volumeYearDiff{} dB(A) reduction between 2020 and 2022 implies a substantial drop in sound energy owing to the logarithmic decibel scale, indicating that everyday exposure may have meaningfully decreased over this period.

First, since high-volume listening was rare and concentrated, safe-listening interventions delivered through earables are most useful when targeted at the specific groups and periods that account for elevated exposure, rather than applied uniformly — positioning the earable as a personalized, exposure-aware hearing-health node rather than a generic volume limiter. Second, and more tentatively: passively logged longitudinal exposure is a prerequisite for such intervention-based studies, because the point measures and self-reports lack the resolution to identify who is at risk and when.


\subsection{Limitations and Constraints}
\label{subsec:limitations}
Our study has several limitations that affect the interpretation and generalizability of its findings.

\subsubsection{Sample and Generalizability} The study's sample was primarily drawn from a German, university-affiliated population that was technologically proficient and willing to share health data from their iPhones. This introduces a sampling bias, potentially excluding groups with different access to technology, privacy concerns, or usage habits. While a survey including non-iPhone users was used to broaden the sample, this selection bias may still limit the applicability of our findings to a broader population. Future research should extend this work to the non-iPhone population as well.

\subsubsection{Data Gaps and Exposure Estimation} Our assumption that periods without logged data represent non-usage likely underestimates total exposure, as participants may have used devices not captured in the iPhone ecosystem. Additionally, the iOS version could have influenced data availability.

\subsubsection{Measurement Precision} The accuracy of Apple's Headphone Audio Exposure metric varies by device. While measurements are highly reliable for Apple and Beats products, which constituted about 68\,\% of our dataset, they are less precise for third-party and wired devices, where Apple Health uses device-specific volume levels. This introduces variability into the absolute exposure values.

\subsubsection{Correlational Findings} Observed relationships between PLD usage and other variables are correlational. The study's design does not allow for determining the direction of causality between these variables.

\subsubsection{Construct Definition} While PLDs and earables are often discussed together, findings related to general PLD usage may not fully capture the specific interaction patterns associated with modern, computationally advanced earables. Also, people with health and accessibility needs might use the devices for extended periods.

\subsubsection{Temporal Overlap with COVID-19 timeline} The study period overlaps with the COVID-19 pandemic timeline, which would have an impact on PLD usage due to increased remote working situations and pandemic stress. Though the yearly trends show an increase in PLD usage in 2022 when the restrictions were relaxed, the results also stabilized until 2024. With just the year as the only predictor, we cannot fully attribute the trend to the pandemic alone without ruling out other confounders.

\subsubsection{Other Research Directions}
Our findings highlight several promising directions for future research in the domain of personal audio technology and hearing health.

\subsubsection{Cross-Cultural and Subgroup Analysis} To establish the broader relevance of our findings, cross-cultural studies are essential. Investigating temporal, social, and demographic patterns across diverse cultural contexts with varying social norms around PLD use will clarify the generalization of our observations. Furthermore, a more granular analysis of user subgroups, such as examining how different facets of sensation seeking relate to physiological arousal and usage patterns, could provide a more nuanced understanding beyond treating high sensation seeking as a monolithic trait.

\subsubsection{Including Other Variables}
Future research should explore PLD usage habits in specific contexts, such as during exercise, and investigate links to physiological factors and hearing health outcomes. Additionally, consider the impact of environmental noise levels on PLD usage.


\section{Conclusion}
This work characterized everyday PLD use, combining survey responses from \participants{} adults with multi-year PLD usage logs from \iPhoneParticipants{} of them. Our results show that logged usage was intermittent and temporally structured, with no listening recorded on \zeroDayRate{}\,\% of participant-days in 2024, and mean daily use was higher in recent years; sensation seeking was associated with heavier self-reported use; participants under 25 listened at higher volumes than the 25--34 group and high-volume exposure was uncommon, though \participantAtleastOneWeekExceed{} participants exceeded the WHO weekly recommendation in at least one week. Most participants also reported avoiding PLD use in social situations.
 
For earable computing, these patterns caution against assuming continuous, uniform wear and instead favor designs aligned with the intermittent windows in which people already wear PLDs, while treating social settings as systematic coverage gaps and safe-listening interventions as most effective when targeted rather than uniform. More broadly, the study shows the value of passively logged, longitudinal data for grounding earable research in realistic behavior, and future work should extend it to more diverse populations and for real-world interventions that manage exposure without undermining the functions PLDs serve.

\begin{acks}
This work is funded by the KIT Center of Health Technologies and supported by the Helmholtz Association Initiative and Networking Fund on the HAICORE@KIT partition.
\end{acks}

\bibliographystyle{ACM-Reference-Format}
\bibliography{acmart}

\newpage
\begin{appendices} \label{sec:rq_detailed}
\section{Research Questions}
\begin{table}[ht]
\renewcommand{\arraystretch}{1.2}
\begin{center}
\scriptsize 
\caption{List of all the research questions and their corresponding specific variables used.}
\begin{tabular}{ p{0.5\textwidth} p{0.1\textwidth} p{0.3\textwidth} } 
 \textbf{Research Question} & \textbf{Source} & \textbf{Data} \\
\toprule
\textbf{Use cases}  & & \\
  \textbf{(RQ1) What are the self-reported everyday contexts of PLD and ANC use?} & &  \\
 1. Which applications are PLDs used for in everyday contexts?? & Survey & 
 1. What do you use headphones for most often? \newline
 2. In which situations do you use headphones most often? \\
 2. How commonly is Active Noise Canceling (ANC) used and for which applications? & Survey &
 1. Do you use noise-canceling/ANC features? \newline
 2. If yes, in which situations do you use ANC? \\
  \textbf{(RQ2) How are PLDs used over time?} &  & \\
 1. Are PLDs used more on weekends than weekdays? & Historical Data  & Historical PLD Usage \\
 2. Is PLD usage increasing over the years?  & Historical Data  & Historical PLD Usage \\
\midrule
\textbf{Users}  & & \\
\textbf{(RQ3) How are psychological traits and social contexts associated with PLD use?} & &\\
 1. Does sensation seeking relate to PLD usage patterns? & Survey & 
 1. BSSS Questionnaire \newline
 2. How often do you use headphones? \newline
 3. How much time per week do you use headphones? \newline
 4. How many hours per week for music with headphones? \\

 2. Do Big Five personality traits relate to PLD usage patterns?  &  Survey & 
 1. Big Five Questionnaire \newline
 2. How often do you use headphones? \newline
 3. How much time per week do you use headphones? \newline
 4. How many hours per week for music with headphones? \\

 3. What are the perceptions and overall attitudes towards using PLDs in social situations? & Survey & 
 1. Do you feel socially isolated by headphones? \newline
 2. Do you feel shielded by headphones? \newline
 3. How do you feel about using headphones in social situations? \\

    \textbf{(RQ4) How do demographics relate to PLD usage?} & &\\
 1. Does one gender group use headphones longer or louder than others? & Survey, Historical Data & 1. Gender \newline 2. Historical PLD Usage \\
 2. Does one age group use headphones longer or louder than others? & Survey, Historical Data & 1. Age \newline 2. Historical PLD Usage \\

  \textbf{(RQ5) How loud are people listening, and is it within WHO guidelines?} && \\
 1. Is PLD listening volume increasing over the years??  & Historical Data & Historical PLD Usage \\
 2. How are listening sessions distributed across sound exposure categories? & Historical Data & Historical PLD Usage \\
 3. Do participants listen within the WHO safe listening recommendation?  & Historical Data & Historical PLD Usage \\

\end{tabular}
\end{center}
\label{tab:rqs-full}
\end{table}
\newpage
\section{Crosstab of social isolation and shielding responses while using PLDs} \label{sec:agreement}

\begin{figure}[ht!]
\centering
\includegraphics[width=\linewidth]{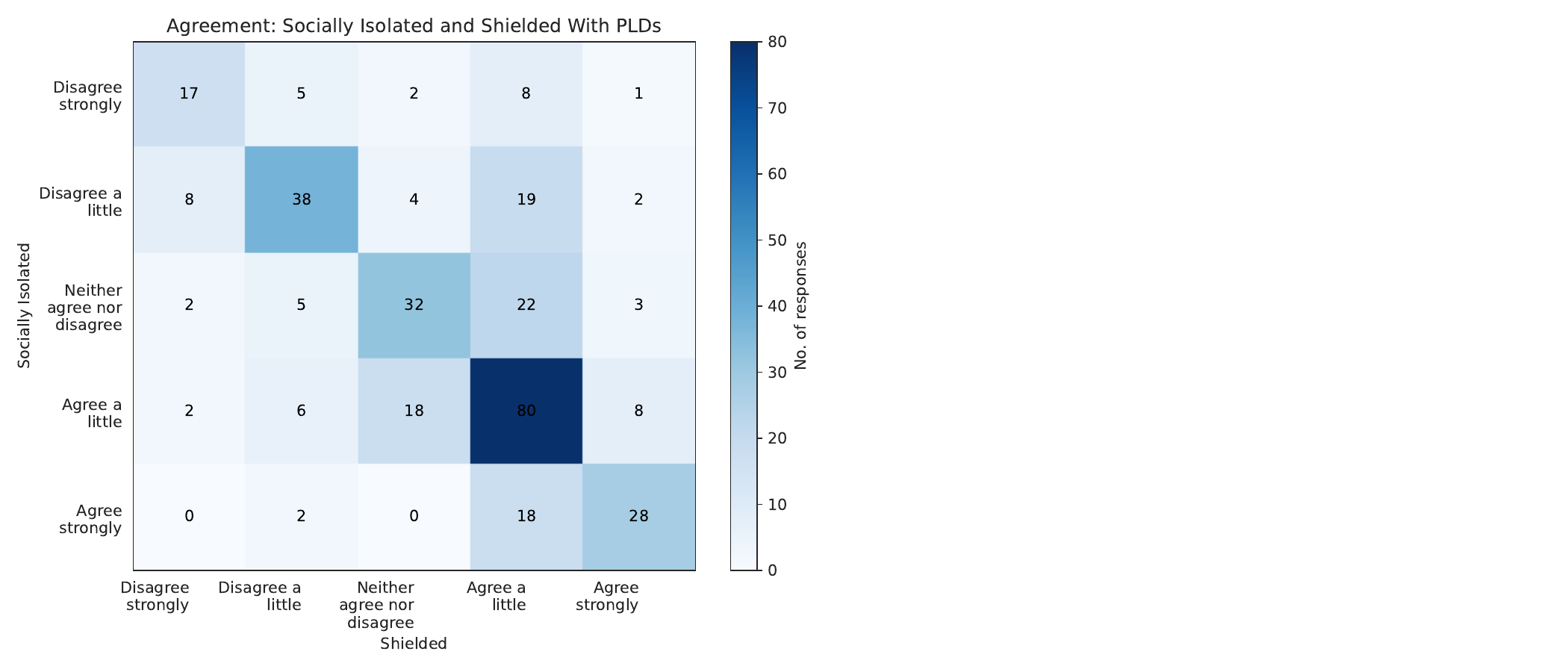}
\caption{Agreement between feelings of being socially isolated and socially shielded when using PLDs, shown as a cross‑tabulation of Likert responses. The feelings of shielding and isolation while using PLDs tend to move together.}
\Description
[Agreement between feelings of being socially isolated and socially shielded when using PLDs.]{Heatmap showing agreement between two 5-point Likert items for two constructs: feeling Socially Isolated (y-axis) and Socially Shielded (x-axis) while using PLDs, from 330 participants. Cell color encodes response count on a white-to-dark-blue scale (0–80). The diagonal — representing agreement between the two constructs — dominates: 17, 38, 32, 60, and 28 participants gave matching ratings, with "Agree a little" on both scales being the most common response (60 participants, darkest cell). Off-diagonal values are small and close to the diagonal, indicating that feelings of social isolation and shielding while using PLDs are positively correlated and tend to move together. }
\label{fig:psy_crosstab}
\end{figure}

\end{appendices}

\end{document}